\documentclass[final,onecolumn]{elsart3p}

\usepackage{amsmath,graphicx,fix2col}

\newcommand{\bez}{B\'ezier}

\begin{document}

\begin{frontmatter}
  
%%%%%%%%%%%%%%%%%%%%%%%%%%%%%%%%%%%%%%%%%%%%%%%%%%%%%%%%%%%%%%%%%%%%%%%%%%%%%%

\title{X-Ray Propagation in Tapered Waveguides: Simulation and Optimization}
  %and Tilted Zone Plates} 

%%%%%%%%%%%%%%%%%%%%%%%%%%%%%%%%%%%%%%%%%%%%%%%%%%%%%%%%%%%%%%%%%%%%%%%%%%%%%%

\author[irp]{Sebastian Panknin},
  \author[ol]{Alexander K. Hartmann}
  and \author[irp]{Tim Salditt}
  \address[irp]{Institut f\"ur R\"ontgenphysik,Universit\"at G\"ottingen,\\
    Friedrich-Hund-Platz~1, 37077 G\"ottingen, Germany}
  \address[ol]{Institut f\"ur Physik, Universit\"at Oldenburg,\\
    26111 Oldenburg, Germany}

\begin{abstract}
We use the parabolic wave equation to study the propagation of x-rays in 
tapered waveguides by numercial simulation and optimization. 
The goal of the study is
to elucidate how beam concentration can be best achieved in 
x-ray optical nanostructures. Such optimized waveguides can e.g.\ be used
to investigate single biomolecules. 
Here, we compare tapering geometries, which can be parametrized by linear
and third-order (B\'ezier-type) functions and can be fabricated using
standard e-beam litography units. These geometries can be described in  two and four-dimensional parameter 
spaces, respectively. In both geometries, we observe
 a rugged structure of the
optimization problem's ``gain landscape''. Thus, the optimization
of x-ray nanostructures in general will be
 a highly nontrivial optimization problem.
\end{abstract}

\end{frontmatter} 

%-----------------------------------------------------------------------------
% \section{Introduction}
%-----------------------------------------------------------------------------

\hyphenation{wave-guide wave-guides}

%%%%%%%%%%%%%%%%%%%%%%%%%%%%%%%%%%%%%%%%%%%%%%%%%%%%%%%%%%%%%%%%%%%%%%%%%%%%%%
\section{Introduction}
%%%%%%%%%%%%%%%%%%%%%%%%%%%%%%%%%%%%%%%%%%%%%%%%%%%%%%%%%%%%%%%%%%%%%%%%%%%%%%

X-ray waveguides are promising novel optical devices to generate x-ray
quasi-point sources for hard x-ray imaging
\cite{DiFonzo2000,Fuhse2006_PRL}.  X-ray waveguides (WG) have been
designed and fabricated as planar layered systems (1D-WG)
\cite{Spiller1973,FengSinha1993,Jark2001} or  as lithographic channels
(2D-WG) \cite{Pfeiffer2002,Jarre2005}.  In both cases (guiding layer
or guiding channel)  mainly waveguides with a core of constant cross
section have been considered.  Beam propagation in such structures,
which are translationally invariant along the propagation axis $z$, is
by now well studied and understood
\cite{Zwanenburg1999,DeCaro2003,Fuhse2006_AO,Fuhse2006_OC}.  As
optical elements, such waveguides serve mainly two purposes:  Firstly,
the radiation modes are damped out and the cross section of the exit
beam is  thus reduced significantly with respect to the incident
beam. Secondly, in the case of single-mode waveguides
\cite{Fuhse2004_APL}, the phase and amplitude of the wave is
determined exactly, and therfore presents an ideal  wave for coherent
imaging. In other words, the waveguide acts as a spatial and coherence
filter, but not to collimate or to concentrate the beam.  The flux
density necessary for nanometer-sized beam with high enough
intensities for practical applications must then be achieved by
focusing elements placed in front of the waveguide, such as
Kirkpatrik-Baez (KB) mirrors \cite{Kirkpatrick1948,Jarre2005}, Fresnel zone
plates \cite{Chao2005} or compound refractive lenses
\cite{crl,schroer}. 
%Such beams are needed for phase contrast projection microscopy of 
%single biological cells, organelles or functional biomolecular complexes. 

Here we address the question to which extent tapering of the waveguide
can be used to concentrate the beam and to filter it at the same time.
This would make pre-focusing optics in front of the waveguide obsolete
for some applications.  A first experimental realization was
demonstrated by tapered air-gap waveguides \cite{Zwanenburg2000b}, but
was restricted to the multi-modal regime. Bergemann et al.\@ have then
studied linearly tapered x-ray waveguides by analytical and numerical
calculation \cite{bergemann}, concentrating on minimum beam width at
the exit. Here we want to generalize the numerical studies to
non-linear tapering profiles. Rather than the minimum achievable beam
width, which has been addressed before \cite{bergemann}, we address
the maximum flux density enhancement as function of tapering
parameters,  and the associated question of an optimum interface
geometry.   As in all problems of reflective optics, e.g.\  bent
mirrors, one would assume that the shape function is a crucial parameter
in the problem.  As a first step, we consider both linear and third
order polynominal interface geometries. Note that the
parameters studied correspond well to structures which can also be
fabricated by e-beam lithography.   The propagation in the stuctures
is studied by numerical solution of the parabolic wave equation (PWE),
as used previously  for standard waveguides
\cite{Fuhse_8SXNS,Fuhse2006_AO}. The use of the PWE will be briefly
explained in the next section, followed by the  simulation results.

%%%%%%%%%%%%%%%%%%%%%%%%%%%%%%%%%%%%%%%%%%%%%%%%%%%%%%%%%%%%%%%%%%%%%%%%%%%%%%
\section{The simulation method}
%%%%%%%%%%%%%%%%%%%%%%%%%%%%%%%%%%%%%%%%%%%%%%%%%%%%%%%%%%%%%%%%%%%%%%%%%%%%%%

The propagation of monochromatic x-rays in materials is   described by
the Helmholtz equation
\begin{equation}
  \label{eq:he}
  [\triangle+k^2n^2]\Psi=0  ~.
\end{equation}
$\Psi$ denotes an electric field component in this scalar wave
equation, $k=\frac{2\pi}{\lambda}$ the wavenumber, and $\lambda$ the
wavelength.  The index of refraction is given by
$n=1-\delta-\mathrm{i}\beta$, where the imaginary part accounts for
absorption. The Laplacian is written as
$\triangle=(\partial_x^2+\partial_z^2)$ for the two-dimensional case
which is of interest here.  The direct solution of this elliptical
differential equation is problematic due to numerical instability and
the fact that the region of interest is large compared to the
wavelength (micrometers up to millimeters compared to
\AA{}ngstr\"om). A numerically much more stable alternative is
provided by  the parabolic wave equation, which can be used in
controlled and excellent approximation  for forward propagation
problems in the x-ray regime \cite{popovbuch,fuhsenat,fuhsefd}.  To
this end, a function $u(x,z)$ is defined by
$\Psi(x,z)=:u(x,z)\exp(-\mathrm{i}kz)$. Neglecting terms of order
$\mathcal{O}(\partial_z^2u)$ which is justified if the $z$-axis  is
almost parallel to the wavevector, the Helmholtz equation is
fullfilled for $\Psi$, under the condition that $u$ solves the
parabolic wave equation
\begin{equation}
  \label{eq:pwe}
  [-2\mathrm{i}k\partial_z+\partial^2_x+k^2(n^2-1)]u=0
\end{equation}
Mathematically, this equation is of the same form as the stationary
Schr\" odinger equation for a massive particle (without spin) in a
potential. After renaming of variables, the potential  in the Schr\"
odinger equation is equivalent to the  the refraction index $n$
profile in the optical case considered here.

In order to simplify the expressions and to find a suitable length
scale for numerical calculations, we use the natural units
$\frac{1}{\delta k}$ for the distance  along the propagation
direction, and $W=\frac{\pi}{k\sqrt{2\delta}}$ for the lateral
dimension, respectively. Note that $W$ is the   critical width at
which a waveguide becomes mono-modal, i.e.\@ the width at which the
waveguide  supports only a single mode
\cite{bergemann,opticwgbuch}. At the same time, a waveguide of width
$d=W$  leads to the highest possible wave confinement. The intensity
distribution of the wave broadens both for smaller and  larger $d$. While the
second case is trivial, it is important to note  that the nearfield
intensity distribution is broadened by evanescent waves in the
cladding, if $d$ is reduced below $W$. Using these natural units
\begin{equation}
  \label{eq:natunits}
  Z:=\delta k z,\qquad X:=\frac{x}{W} ~,
\end{equation}
 \eqref{eq:pwe} can be rewritten as
\begin{equation}
  \label{eq:pwenat}
  \partial_Zu=-\frac{\mathrm{i}}{\pi^2}\partial_X^2u
             +\left\{\begin{array}{ll}
             (i-\frac{\beta}{\delta})u,&\quad \text{in material}\\
             0,&\quad \text{in vacuum}.
             \end{array}\right.
\end{equation}
The ratio $\frac{\beta}{\delta}$ remains the only material dependent
parameter in this equation.

The equation is solved using a Crank-Nicolson finite-difference
scheme, which was already proved to give reliable results for x-ray
waveguides and Fresnel zone plates \cite{fuhsefd,popovpwg}. For the
initial values, a plane wave propagating along $z$ was assumed. The
lateral boundary conditions are given by a damped plane wave
propagating in the material far away from the region of interest.

For field propagation in vacuum, the  Fresnel-Kirchhoff integral 
for two-dimensional beam propagation \cite{zylinderfk} was
used \cite{bornwolf},
\begin{equation}
  \label{eq:fk}
  \Psi(x,z)=\frac{\mathrm{i}}{\sqrt{\lambda}}
                \int\mathrm{d}x'\Psi(x',0)
                \frac{\mathrm{e}^{-\mathrm{i}kr}}{\sqrt{r}}\cos\alpha
                ~,
\end{equation}
where $\alpha$ is the angle between the $z$-axis and the vector
$(x-x',z)$. The corresponding distance to the origin is
$r^2=z^2+(x-x')^2$.

%%%%%%%%%%%%%%%%%%%%%%%%%%%%%%%%%%%%%%%%%%%%%%%%%%%%%%%%%%%%%%%%%%%%%%%%%%%%%%
\section{Tapered x-ray waveguides: shape optimization}
%%%%%%%%%%%%%%%%%%%%%%%%%%%%%%%%%%%%%%%%%%%%%%%%%%%%%%%%%%%%%%%%%%%%%%%%%%%%%%

We have considered two generic types of tapering geometry: (i) the
linear taper, and (ii) a tapering function parameterized by B\' ezier
curves.  For comparison with experimental values,  we have chosen
material parameters and photon energy $E$ according to recent
experiments and present fabrication methods.  Vaccuum channels in $Si$
have been simulated at a photon energy $E=12\,\mathrm{keV}$. Taking
air/vaccuum as the guiding material simplifies simulations because
only the index of refraction of the cladding differs from
unity. Optically, the structures are still very close to the
experiments with polymer channels in $Si$ \cite{Jarre2005,fuhsediss},
and are completely equivalent to the  second generation of x-ray
waveguide channels fabricated by e-beam lithography, ion etching and
subsequent wafer bonding of a cap wafer.  For both tapering
geometries, the waveguide end width was held constant at
$d=1,\,W=20\,\mathrm{nm}$ which is the critical width
$W=\frac{\pi}{k\sqrt{2\delta}}$ for single-mode waveguides
\cite{bergemann,opticwgbuch}. At this value, the resolution limit of
waveguides is reached, corresponding to a full width at half maximum
$\mathrm{FWHM}_\mathrm{min}=0.64\,W$ for the intensity profile in the
channel of width $W$ \cite{bergemann}.

\begin{figure}[htbp]
  \centering \includegraphics[width=\linewidth]{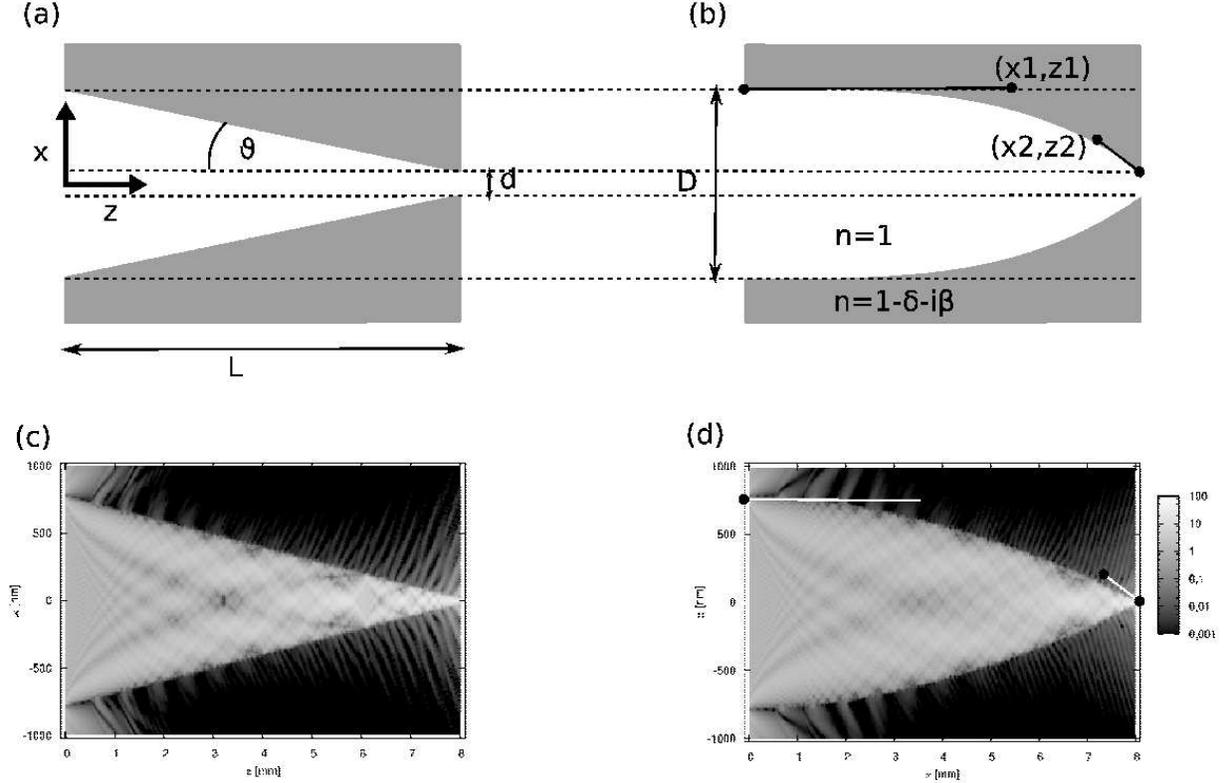}
  \caption{Shape functions and field distribution of taperd
  waveguides: the top row shows the schematics, the bottom row the
  calculated intensity distribution for the optimal parameters of each
  kind. The waveguides consists of vacuum channels embedded in Silicon
  at a photon energy of $E=12\,\mathrm{keV}$.  (bottom left) Linear
  taper with an opening angle $\vartheta=0.037\,\vartheta_c$,
  transporting radiation  collected within a width of $D=0.77 \mu$m
  over a propagation distance of $L=8\,\mathrm{mm}$  down to the
  critical exit cross section of $d=19.88\,\mathrm{nm}$, corresponding
  to the critical width $W$ of $Si$. (bottom right) Waveguide  tapered
  with a \bez curve interface. The lengths $d$, $D$ and $\ell$ as
  above. The two control vectors are $P_1=(x_1=3.45\,\mathrm{mm},
  y_1=766.2\,\mathrm{nm})$ and
  $P_2=(x_2=7.25\,\mathrm{mm},y_2=176.7\,\mathrm{nm})$, respectively }
  \label{fig:wg:types}
\end{figure}

The goal of the optimization was to quantify the field propagation and
confinement, for given opening width
$D$ and end width $d$ or for given $d$ only. 
In general, one can expect that optimization of
x-ray nanostructures might be highly non-trivial, as for many
optimization problems occurring in physics \cite{opt-phys2001,opt-phys2004}.
Nevertheless, since we restricted ourself here to the two above mentioned
tapering  geometries, we could use a quite simple optimization algorithm,
see below. The two tapering
geometries were parameterized as follows (see
Fig. \ref{fig:wg:types}):
\begin{itemize}
\item In the linear case, a straight interface seperates the channel
  from the cladding at an opening angle $\vartheta$. The length of the
  waveguide is denoted by $L$.
\item In the case of a B\'ezier curve interface, the shape is
  controlled by two vectors $\mathbf{P}_1=(x1,y1)$ and
  $\mathbf{P}_2=(x2,y2)$, which determine the derivative of the curve
  at the endpoints and thereby fully define a curve of third
  polynomial degree \cite{Bronstein}.
\end{itemize}
We have used these functions also because they can be handled by
pattern generators of modern e-beam lithography units.

\begin{figure}[htbp]
  \centering

%\remarkAlex{$l$ $\to$ $L$ im Bild}

  \includegraphics[width=0.9\textwidth]{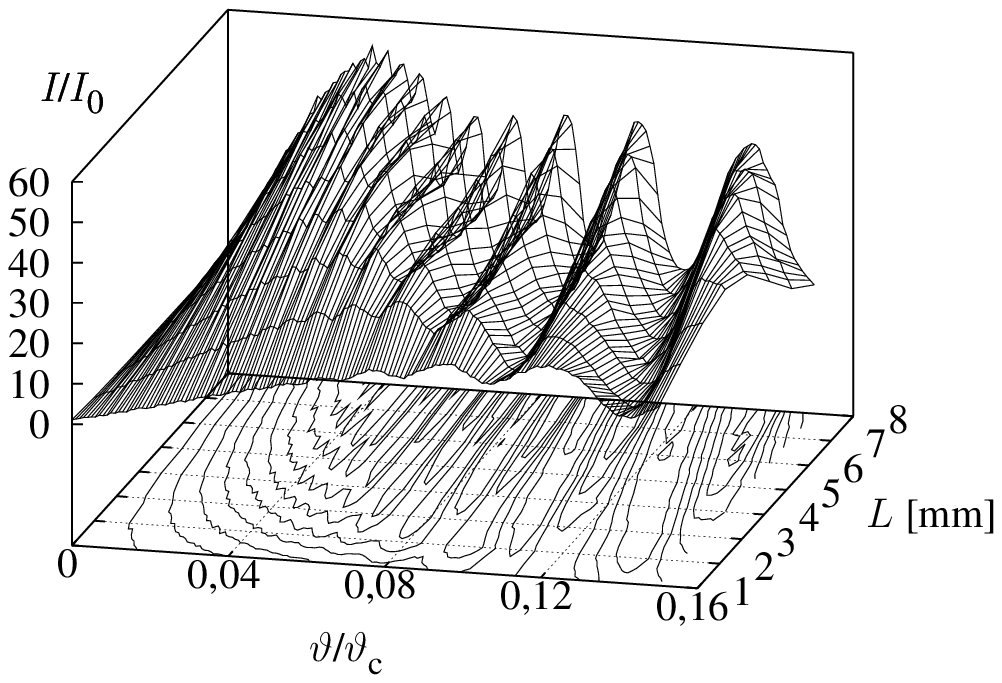}

  \includegraphics[width=0.8\textwidth]{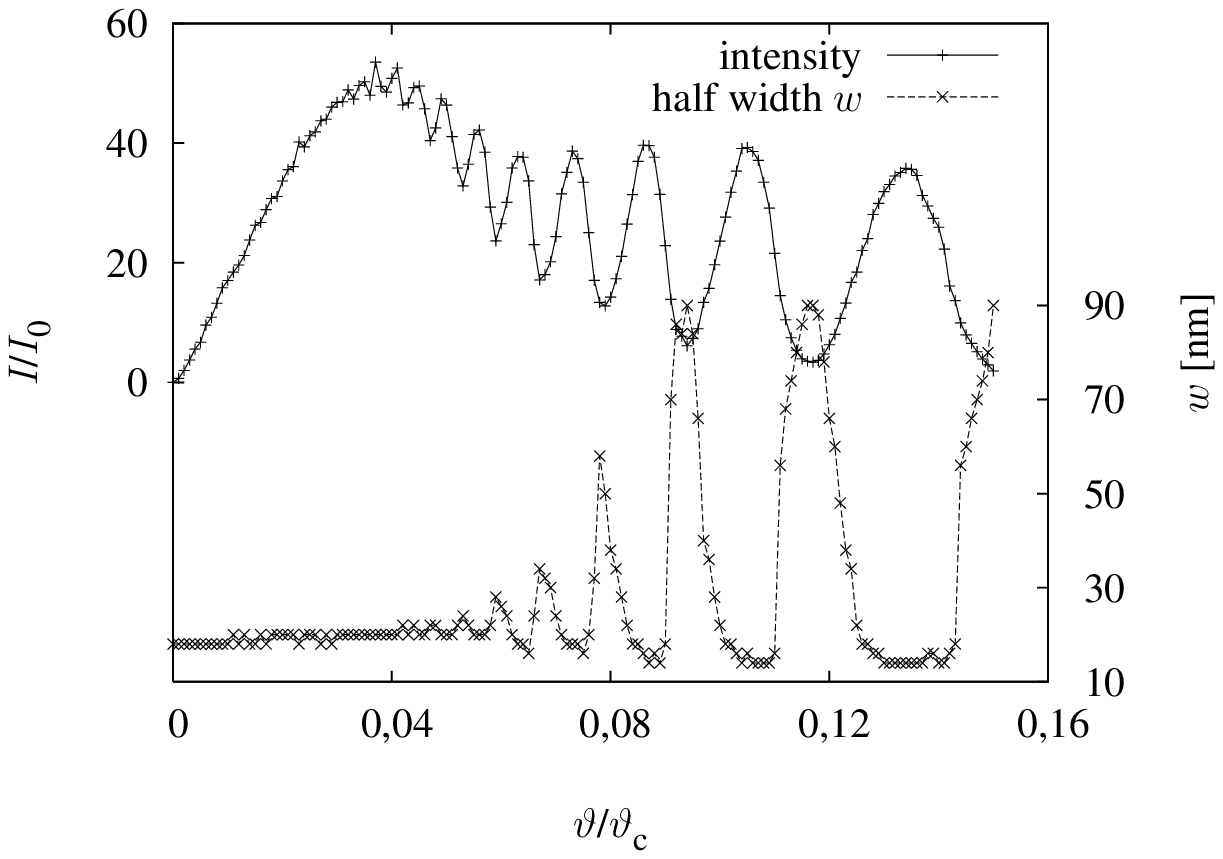}

  \caption{(top) Maximum intensity $I(\vartheta, L)$ at the exit
  normalized to the incident intensity $I_0$, as a function of opening
  angle $\vartheta$ and waveguide length $L$ for linear taper. (bottom)  Maximum
  intensity $I(\vartheta, L= 8{\rm mm})$ (left axis) and beam width
  FWHM (right axis) at the waveguide exit, as a function of
  $\vartheta$. The oscillations are phase shifted with respect to each
  other.
  \label{fig:wg:ltheta}}
\end{figure}

Firstly, let us consider the linear taper. Waveguides with linear
taper have been simulated for different length $L$ and angle
$\vartheta$.  Fig. \ref{fig:wg:ltheta} shows the resulting peak
intensities $I$ normalized to the incident intensity $I_0$, i.e. the gain
$I/I_0$,
at the end of the waveguide (exit plane)  as a
function of $L$ and $\vartheta$. Hence, this plot displays
 the ``gain landscape''
of the optimization problem. Even for this two-parameter space,
a very rugged gain landscape is visible:  
With increasing opening width $D
\simeq 2 L \tan(\vartheta)$, $I$ first increases,  in proportion to
the geometric acceptance. The initial increase of $I(\vartheta, L= {\rm
const})$ crosses over to a surprisingly broad range of angles where
the radiation transport and collimation raimains high, but is
modulated by an oscillation occuring with a periodicity at which
small. Finally, the intensity decreases when $\vartheta$ becomes too
large, leading to a leaking of the intensity out of the confining
interfaces. This leakage occurs for multireflections of high order
$n$. The condition  to prevent the leakage is simply given by
$n\vartheta\le\vartheta_c$, where $\vartheta_c$ is the angle of total reflection,
  as expected from ray optics.
Nevertheless, at the end of the waveguide a dicrete set of rod-like
intensity spikes are observed in the simulation, travering or
'leaking' from the interface, see again Fig. \ref{fig:wg:types}.
 These rods or spikes originate from
interference effects which would correspond to transmitted beams after
multiple reflections in the ray-optical description. When $\vartheta$
is varied, these rods move and give rise to an oscillatory behavior in
the exit plane. This effect is characterized by the oscillation
between  either sharp and intense profile at the exit, i. e. high $I$
and small with (full width at half intensity FWHM) or a smeared
profile with low $I$ and large FWHM.  In summary, $I(\vartheta ,L=
{\rm const})$ possesses an optimum angle $\vartheta$, while
$I(\vartheta={\rm const}, L )$ increases with the length $L$, if
$\vartheta$ is chosen accordingly, at least over the simulated range
in $L$. We chose $L=8\,\mathrm{mm}$ as an experimentally achievable
upper limit, see Fig. \ref{fig:wg:ltheta}.  For this length,
the optimal angle is $\vartheta \simeq 0.037 \vartheta_c$ leading to
an opening width of $D=0.77\mu$m and $I/I_0 = 51.4$.

Now let us consider tapering of the B\'ezier type. Obviously, more
parameters are needed to describe the  space of  B\'ezier curves which
includes the linear as well as the parabolic case.  Correspondingly,
an optimization of the parameters is needed, since the optimum can no
longer be  found 'by eye' from running a set of simulations. As a
first step to a complete optimization, we optimized curves by
simulating waveguides for different $y_1$ and $y_2$ at fixed
$x_1=\frac{1}{3}L$ and $x_2=\frac{2}{3}L$. 
As in the linear case the exit was set to the
critical width $d=W$. $D$ and $L$ were kept constant.

\begin{figure}[htbp]
  \centering \includegraphics{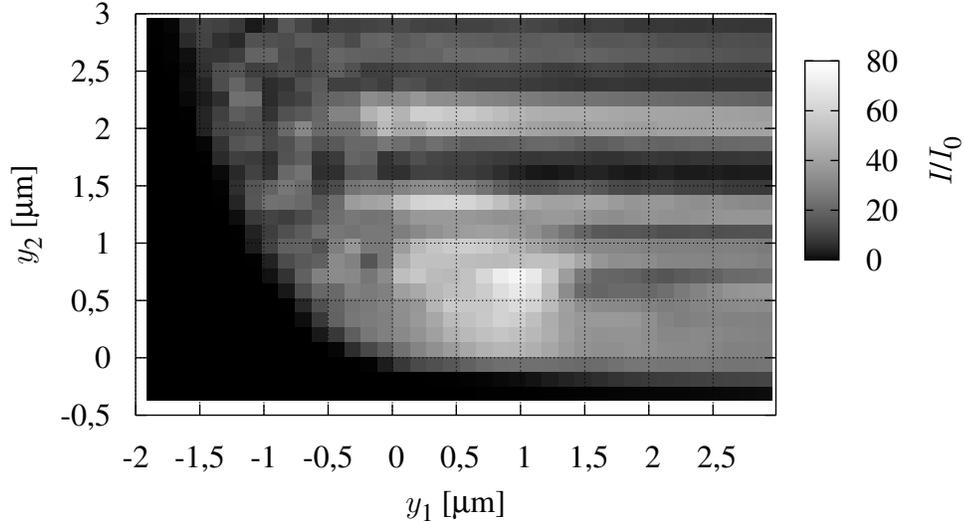}
  \caption{Maximum intensity $I$ (normalized by $I_0$) of a waveguide with B\'
  ezier-type tapering  parameterized by $P_1 = ( 1/3 L, y_1)$ und $P_2
  =(2/3 L, y_2)$.  The function $I(y_1,y_2)$ exhibits a well
  pronounced global maximum.}
  \label{fig:wg:y1y2}
\end{figure}

%We proceeded as follows. First the two-dimensional parameter space
%$y_1, y_2$ was covered by simulations  carried out on a rather coarse
%grid or parameters. 
The results, see Fig. \ref{fig:wg:y1y2}
showed again a rugged gain landscape with a couple of local
minima. Nevertheless,  the maximum intensity seems to occur close to
the linear case, and  a few local minima are present in this
particular 
region. Therefore, a simple approach was feasible: using the amoebae
algorithm \cite{numrec}, the optimum parameters $P_1$, $P_2$ were
searched,  starting from the solution of the linear case $P_1=(\frac{1}{3}L,
\frac{D}{2} - \frac{1}{3}A)$ and $P_2=(\frac{2}{3}L, \frac{D}{2} -
\frac{2}{3}A)$ with $A=\frac{1}{2}(D-d)$.  Note that using other starting
values, the algorithm also found different local maxima,
proving the complexity of the underlying gain landscape and the usefullness
of our pre-scanning of the parameter space.
  As expected,
the maximum intensity is higher for the more general B\'ezier case
than for the linear taper, namely $82.2$ versus  $51.4$  for the peak
gain $I/I_0$, and $F=72.24$ versus $63.27$ for the integrated
flux, see Tab. \ref{tab:optwg}.   However, this difference is
surprisingly small.
%Also some curves are found which effectively increase the opening width
%$D$ and shall so be ignored. 
%$\emph{Maybe some text about middle junctions.}
The result are tabulated in Tab. \ref{tab:optwg}, showing  maximum
intensity $F$, integrated flux $J=\int \mathrm{d}x I$, width $w$
(FWHM) of intensity in the exit plane, as well as the angular
acceptance on the incidence side, defined as the width (FWHM) of
$I(\alpha)$,  where $\alpha$ is the incidence angle $\alpha$. Note
that if not otherwise stated, the angle $\alpha$ between
incident beam and waguide axis was kept constant at
$\alpha=0$. In addition to the two tapering geometries,  the planar
waveguide is included as a reference.  As expected, the angular
acceptance $\omega$ is much smaller for the tapered waveguides due to
conservation of photon density in phase space (brilliance).
%Further the B\'ezier borderline optimizes the maximal intensity 
%by smaller outgoing beam width and
%so the flux is not much improved.

\begin{table}[tb]
  \centering
  \caption{Simulation results: maximum intensity $F$, flux $J$, width
    (FWHM) of intensity at the exit $w$, and FWHM for the angular
    dependency of the intensity (angular acceptance) $\omega$ are
    tabulated for (a) the plane, (b) linear tapered and (c) a
    waveguide with interfaces parameterized as B\'ezier curves
    (values obtained after parameter optimization).}
  \label{tab:optwg}
  \vspace{12pt}
  \begin{tabular}{lrrr}
    &(a) planar&(b) linear &(c) B\'ezier\\ \hline \rule{0 pt}{14
    pt}intensity $\frac{I}{I_0}$ &$11,1\cdot10^{-3}$&$51.4$&$82,2$\\
    \rule{0 pt}{14 pt}flux $\frac{J}{I_0
    d}$&$10,4\cdot10^{-3}$&63,27&72,24\\ \rule{0 pt}{14 pt}beam width
    $w$&15,2&17,8&12,0\\ \rule{0 pt}{14 pt} angular acceptance
    $\frac{\omega}{\vartheta_c}$& 0,929&0,031&0,025
  \end{tabular}
\end{table}

\section{Summary and Conclusion}
%%%%%%%%%%%%%%%%%%%%%%%%%%%%%%%%%%%%%%%%%%%%%%%%%%%%%%%%%%%%%%%%%%%%%%%%%%%%%%

We have studied beam propagation and concentation in tapered x-ray
waveguides by simulations of finite-difference equations and subsequent
parameter screening and optimization. The local slope of the waveguide
interfaces and  the associated tapering geometry is limited by the angle
of total reflection, which restricts the tapering geometry to very
elongated shapes. Over a device length on the order of $10$mm, an
incoming beam with a width of   $0.5-1\mu$m can be concentrated to the
minimum width (FWHM) $w= 0.65 W \simeq 13 \,\mathrm{nm}$ for
silicon. The associated expected flux enhancement in the range of
$40-80$ for one-dimensional focusing is observed in the simulation,
confirming that most of the radiation is indeed transported to the
exit aperture.  For two-dimensional focusing, the gain would be
squared, and would thus range up to about $5\times 10^3$, if $10$mm long
structures are available. Note that both lithographic wafer processing
and hollow glass fibers with well controlled shapes  are conceivable
even on length scales of $100$mm.  However, let us stay withing a more
conservative estimate based on a $10$mm long waveguide. As a result,
the flux density gain, the size of the entrance aperture, and the
angular acceptance $\omega$ are fully compatible with an optical
scheme, where a synchrotron source is imaged by a lense of moderate
numerical aperture (mirror system or CRL) onto the entrance aperture
of a tapered waveguide. Let us consider realistic values for 3rd
generation synchrotron sources: a source on the order of  $100\mu$m
cross section is demagnified by CRL or KB optics of focal length
$f\simeq 1-5$m positioned $100$m behind the source.  In this case, the
focal spot and the convergence angle of the prefocussing optics can be
matched  to the entrance aperture and the angular acceptance $\omega$
of the tapered waveguide, respectively.  In this case the combined
gain of the pre-focussing optics and the tapered waveguide would lead
to an intense and collimated virtual source, which is unlikely to be
achieved by a single optical device.  Most importantly, the simulation
results show that the performance of the tapered waveguide does not
depend on the shape function in a very sensitive way. Indeed, the beam
can be funneled to the exit aperture surprisingly well already with a
linear taper. In sharp contrast to single bounce mirror optics, the
tapered waveguide almost forces the convergence of the beam, and is
thus much less sensitive to figure errors.
Nevertheless, we have also observed a rather rugged gain landspace for this
optimization problem. Since the optimization was only for a
 few-parameter space, a simple algorithm could be used here.
We expect that for the optimization of
more  complex x-ray nanostructures,
like layered materials, more sophisticated algorithms have to be applied.

%%%%%%%%%%%%%%%%%%%%%%%%%%%%%%%%%%%%%%%%%%%%%%%%%%%%%%%%%%%%%%%%%%%%%%%%%%%%%%
\section*{Acknowledgment}
%%%%%%%%%%%%%%%%%%%%%%%%%%%%%%%%%%%%%%%%%%%%%%%%%%%%%%%%%%%%%%%%%%%%%%%%%%%%%%

We thank Christian Fuhse for helpful discussions and the foundation
work concerning the finite difference equations simulations in our group. We gratefully
acknowledge funding by Deutsche Forschungsgemeinschaft (DFG) through
Sonderforschungsbereich 755 {\it Nanoscale Photonic Imaging}.

\bibliography{lit2,literature}
%\bibliography{literature}

\end{document}